\documentclass{PoS}

\usepackage{multicol}
\def\ergscm{erg~s$^{-1}$~cm$^{-2}$}

\def\apj{ApJ}

\def\aap{A\&A}
\def\mnras{MNRAS}

\def\procspie{\ref@jnl{Proc.~SPIE}}   

\def\x1{NGC~5643~X-1}
\def\apec{APEC}
\def\nu{\textit{NuSTAR}}

\title{NuSTAR observation of the Arches cluster: X-ray spectrum extraction from a 2D image}

\ShortTitle{NuSTAR observation of the Arches cluster in 2015}

\author{\speaker{Roman Krivonos}\\
      Space Research Institute (IKI), Moscow, Russia\\
      E-mail: \email{krivonos@iki.rssi.ru}}

    \abstract{The {\it NuSTAR} mission performed a long (200~ks)
      observation of the Arches stellar cluster in 2015. The emission
      from the cluster represents a mixture of bright thermal
      ($kT\sim$2~keV) X-rays and the extended non-thermal radiation of
      the molecular cloud around the cluster. In this work we describe
      the method used to decouple spatially confused emission of the
      stellar cluster and the molecular cloud in the \nu\ data.}

\FullConference{11th INTEGRAL Conference Gamma-Ray Astrophysics in Multi-Wavelength Perspective\\
		 10-14 October 2016\\
		 Amsterdam, The Netherlands}

\begin{document}

\section{Introduction}

The Arches cluster is a young, densely packed massive star cluster in
our Galaxy that shows a high level of star formation activity. It is
located in the inner Galactic Center (GC) region at the projected
distance of $11'$ from the position of the dynamic center of the
Galaxy -- Sagittarius A*.

The Arches cluster is a known source of thermal and non-thermal X-ray
emission. The thermal emission is thought to originate from multiple
collisions between strong winds of massive stars
\cite{chlebowski1991,zadeh2002,wang2006,capelli11a}, and is localized
within the core of the cluster that is about $9''$ ($\sim0.35$~pc at
8~kpc) in radius \cite{figer1999}. Diffuse non-thermal X-ray emission,
revealed by its bright fluorescent Fe~K$\alpha$~$6.4$~keV line
emission, has been detected from a broad region around the cluster
\cite{wang2006,tsujimoto2007,capelli11b,T12,K14}.

The core of the Arches cluster has been relatively well studied in the
standard $2-10$~keV energy band with the \textit{Chandra} and
\textit{XMM-Newton} observatories. The origin of the non-thermal
extended emission, whether it is produced by the photoionization of
the cloud by X-ray photons or through excitation by CR particles, was
considered in many relevant studies
\cite{capelli11b,T12,K14}. \cite{clavel2014} analysed the long-term
behavior of the Arches cloud over 13 years and reported a 30\%
decrease in Fe K$\alpha$ line and continuum flux of the cloud emission
in 2012-2013, providing the evidence that the majority of the variable
non-thermal emission is due to X-ray reflection. Despite LECR-only
emission is almost excluded based on variability, one could expect
that steady background level is a result of CR heating, while most of
the varying emission is due to reflection. For this reason we continue
to measure spectral shape of the Arches cloud non-thermal emission,
which contains imprints of the emission mechanism.

To separate thermal emission of the Arches cluster and non-thermal
emission of the surrounding molecular cloud with \nu\ \cite{nustar}, we
utilize 2D image analysis as demonstrated in
\cite{krivonos2017}. Similar procedure was also applied by
\cite{krivonos2016}, who analysed \nu\ data of local Seyfert 2 active
galactic nucleus (AGN) NGC~5643 and successfully decoupled partially
confused spectra of AGN core and ultra-luminous source located in the
same galaxy. In this paper we present 2D image spectral extraction
procedure in more details.

\section{The standard approach}
\label{sec:standard}

The canonical method of spectrum extraction from X-ray data implies
selecting circular region around the source and the region with
representative source-free background. Then the standard tools of a
given X-ray mission extract events from the regions; calculate the
exposure; apply corrections for the Point Spread Function (PSF),
vignetting, dead-time, etc.; generate appropriate Response Matrix
Files (RMFs) and Ancillary Response Files (ARFs); and finally assemble
source and background Pulse Height Amplitude (PHA) files with updated
header keywords for spectral modeling in one of the popular packages,
e.g. {\sc xspec} \cite{xspec} or {\sc sherpa} \cite{sherpa}. The
standard approach fails in complicated cases with non-uniform
background or spatial confusion of the sources in the X-ray
images. The spatial modelling both the emission of the sources and
background provides natural solution to this problem, however requires
more sophisticated algorithms.

\section{2D image fitting procedure}
\label{sec:2d}

We first combined the \nu\ data into sky mosaics in 15 energy bands
logarithmically covering the \nu\ working energy range
$3-79$~keV. Each data set contains counts and exposure map. In the
following analysis we assume flat background, so we do not need
non-uniform background map, normally produced as a separate map in the
same pixel resolution. Additional map contains PSF extracted from the
\nu\ Calibration Data Base (CALDB), placed in the center of the
map. We assume the same PSF shape over the considered energy range
$3-79$~keV.

We then constructed spatial model of the Arches cluster complex, which
includes two 2D Gaussians. The first represents the cluster's core
emission, with the position fixed at the centroid coordinates measured
in \cite{K14} and width fixed at $4''$ FWHM (PSF smearing effect). The
second Gaussian was aligned with the corresponding ``halo'' Gaussian
component used in \cite{K14} to describe the extended cloud emission,
setting the FWHM model parameter at $72''.4$. We fixed sky positions
of the Gaussians as described in \cite{krivonos2017}. Thus, the only
amplitudes of the 2D Gaussians were free parameters. As mentioned
above, we assume flat background over the detector image. The
normalization parameter was estimated in the annulus $70''<R<130''$
around the cluster.

We use {\sc sherpa} modeling and fitting package, a part of {\sc ciao}
software \cite{ciao}, to construct 2D models and fit them to
data. {\sc sherpa} enables one to construct a complex models from
simple definitions, e.g. our current model setup can be expressed as
$psf(gauss2d.G1+gauss2d.G2)*emap+const2d.bkg*emap$, where
$psf(gauss2d.G1+gauss2d.G2)*emap$ means convolution of two Gaussians
$G1$ and $G2$ with the PSF multiplied by the exposure map $emap$, the
constant background term $bkg$ is added as $const2d.bkg*emap$. Note
that the background is not convolved with the PSF.

\begin{figure}
\includegraphics[width=\columnwidth]{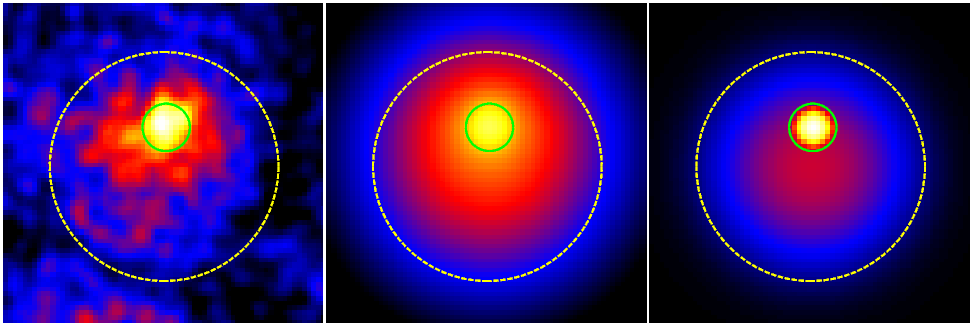}
\caption{An example of 2D image fitting procedure of point-like Arches
  cluster's core thermal emission (2D Gaussian model $G1$, $15'$
  radius green circle) and the extended non-thermal emission of the
  surrounding molecular cloud (2D Gaussian model $G2$, yellow dashed
  $72''.4$ FWHM radius circle). {\it Left}: \nu\ image of the Arches
  cluster complex in $5.8-7.2$~keV band, compatible with fluorescent
  iron line emission of the cloud ($G2$) at 6.4~keV and emission line
  of ionized iron at energy 6.7~keV of the cluster ($G1$). {\it
    Middle}: Best-fit spatial model of the Arches cluster complex in
  $5.8-7.2$~keV band with two gaussians $G1$ and $G2$ convolved with
  the \nu\ PSF. {\it Right}: The same as middle plot, but not
  convolved with the PSF.}\label{fig:2dfit}
\end{figure}

\begin{figure}
\begin{center}
\includegraphics[width=0.7\columnwidth]{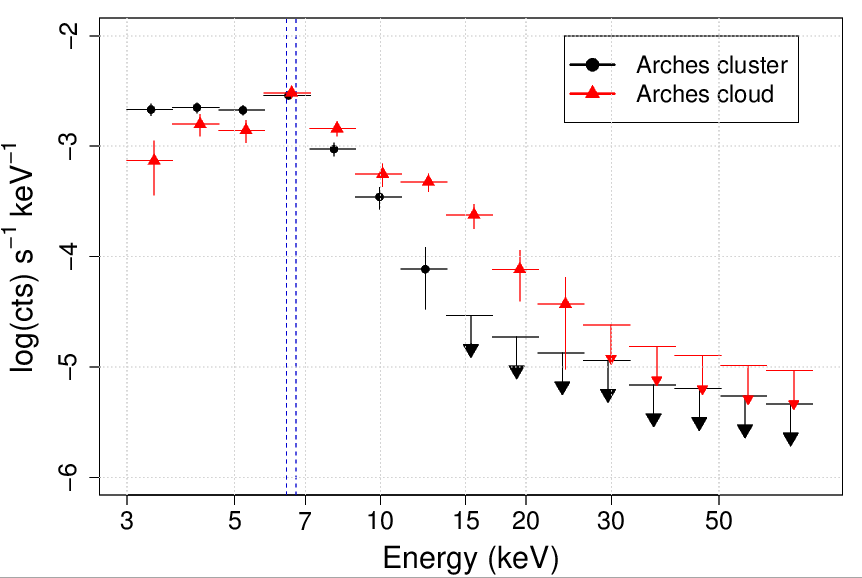}
\caption{Spatially resolved \nu\ spectra of the Arches cluster core
  (black) and cloud (red) X-ray emission. The position of Fe~K$\alpha$
  6.4 and 6.7~keV are marked by vertical dashed lines, as in the
  NuSTAR standard spectra these two lines cannot be separated by our
  technique.  The upper limits are 1$\sigma$
  errors.}\label{fig:2d}
\end{center}
\end{figure}

By running the fitting procedure in each of the 15 energy bands, we
estimated the best-fitting parameters of the spatial models $G1$ and
$G2$. We then calculated the flux, i.e. amount of collected photons
with each model divided by the exposure time. The corresponding errors
were estimated from multi-variate normal distribution, based on
sampling the set of thawed parameters and calculating the model flux
({\it normal\_sample} function in {\sc sherpa}). The estimated model
fluxes and uncertainties were combined into PHA spectra files of each
spatial model $G1$ and $G2$ shown in Fig.~\ref{fig:2d}. The
corresponding Redistribution Matrix File (RMF), which maps from energy
space into PHA space, was simply adopted from the standard spectral
analysis of the `calibration' source PMN~J0641-0320 (see below) with
{\it nuproducts} and rebinned with {\it rbnrmf} tool of {\sc
  HEASOFT}~6.19 package.

As seen from Fig.~\ref{fig:2d}, spatially confused emission of the
Arches cluster complex was effectively decoupled with 2D image fitting
procedure into the soft thermal emission of the cluster and hard
non-thermal emission of the molecular cloud. Since the applied method
uses the full collective power of the PSF, we could detect hard X-ray
emission up to $20-30$~keV. X-ray emission of the cluster contains an
excess in the $5.8-7.2$~keV range, compatible with $\sim6.7$~keV line
and rapidly drops above $\sim10$~keV as expected for thermal emission
with $kT\approx2$~keV. Non-thermal emission of the extended cloud
component apparently includes excess around $6.4$~keV and dominates
above $10$~keV.

To use spatially decoupled $G1$ and $G2$ spectra in {\sc xspec}
spectral modeling package we calibrated effective area (ARF) utilizing
the \nu\ data of a bright source with known spectrum. To this end we
used 20~ks observation (ObsID: 80001003002) of MeV Blazar
PMN~J0641-0320 with very hard power-law spectrum of $\Gamma\approx1$
detectable up to $\sim80$ keV \cite{marco2016}. We have done standard
spectral extraction (Sect.~\ref{sec:standard}) of PMN~J0641-0320
within $70''$ radius circle, and estimated source flux
phot~s$^{-1}$~cm$^{-2}$ in each of the 15 energy bands. We then
repeated 2D image spectral extraction procedure (Sect.~\ref{sec:2d})
for PMN~J0641-0320 as a point-like source (one 2D Gaussian with $4''$
FWHM size) and estimated model flux phot~s$^{-1}$ in each of the 15
energy bands. Comparing the model fluxes obtained by two methods, we
extracted the effective area in cm$^{-2}$ for each band and combined
them into ARF file. We should note that here we assume that the ARF
calibrated for a point-like source is suitable for the extended
emission of the Arches cloud. Given the limited statistics of the \nu\
Arches cluster observations, the deviations are within the
uncertainties.

\begin{figure}
\center
\includegraphics[width=0.9\columnwidth]{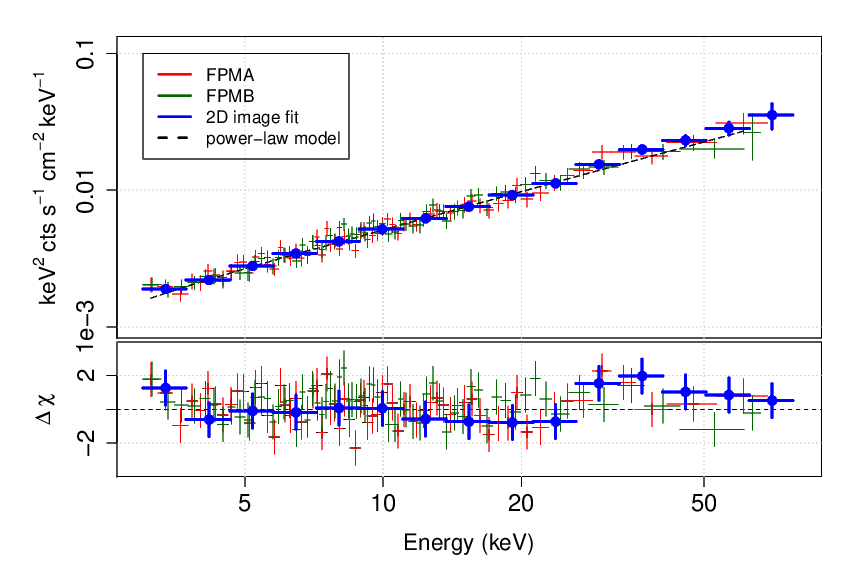}
\caption{\nu\ spectrum of MEV blazar PMN~J0641-0320 extracted from
  $R=70'$ circular region with standard procedures (red and green),
  and spectrum obtained through 2D image analysis (blue) described in
  this paper.}\label{fig:blazar}
\end{figure}

Fig.~\ref{fig:blazar} shows X-ray spectra of MeV Blazar PMN~J0641-0320
obtained with standard method (Sect.~\ref{sec:standard}) for two \nu\
modules FPMA and FPMB, and combined FPMA+FPMB spectrum extracted with
2D image algorithm (Sect.~\ref{sec:2d}). It is seen that
spectra are consistent with each other, demonstrating reliability of
the 2D image fitting approach.

\section{Results}

\begin{figure}
\center
\includegraphics[width=0.49\columnwidth]{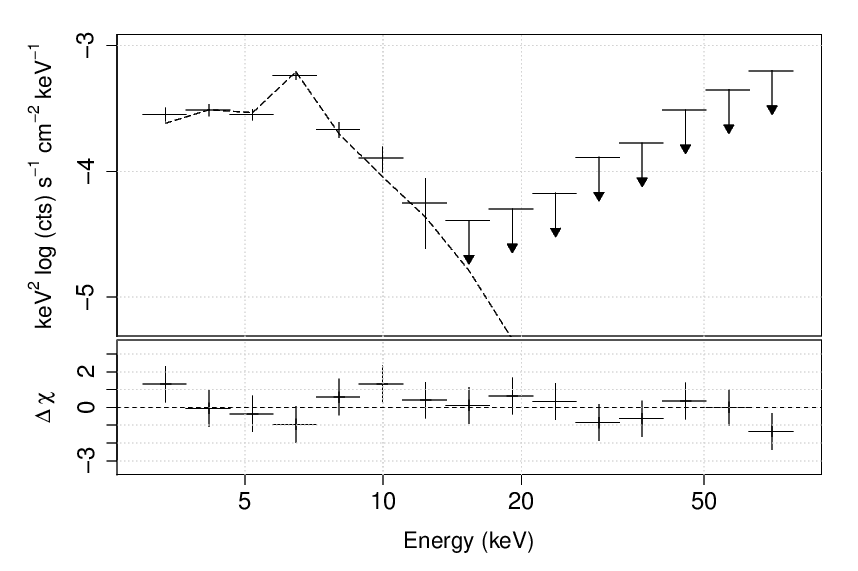}
\includegraphics[width=0.49\columnwidth]{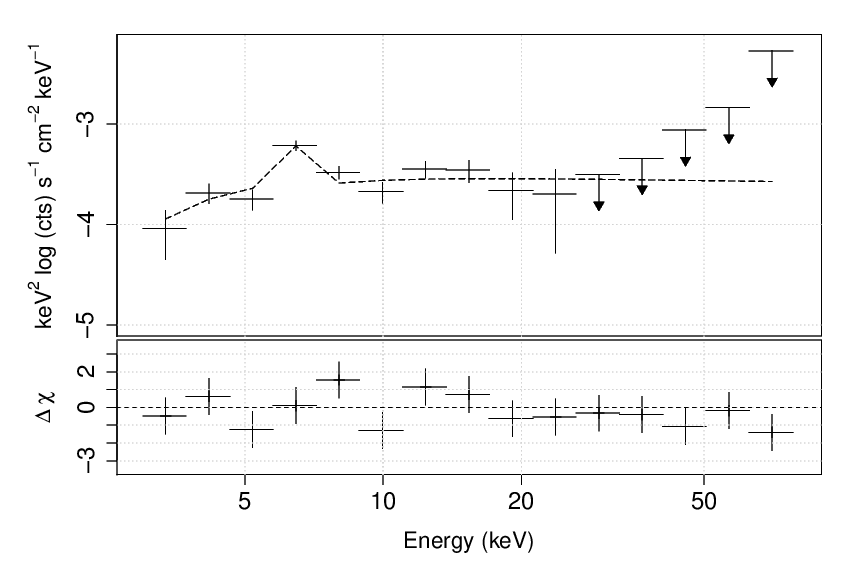}
\caption{Spatially decoupled \nu\ spectrum of the Arches cluster
  complex obtained with 2D image fitting procedure. {\it Left:}
  thermal spectrum of the Arches stellar cluster ({\it APEC}, $kT\sim2$~keV). {\it
    Right:} non-thermal X-ray spectrum of the Arches molecular cloud
  extended emission, approximated with power-law model and iron 6.4
  keV line represented with Gaussian.}\label{fig:res}
\end{figure}

The final spatially decoupled spectra of the Arches stellar cluster
and extended molecular cloud emission is shown in Fig.~\ref{fig:res}.
We fitted the stellar cluster emission spectrum with one \apec\ model
subjects to a line-of-sight photoelectric absorption fixed at $N_{\rm
  H}=9.5\times 10^{22}$~cm$^{-2}$. This simple model provides an
acceptable fit to the data with $\chi^{2}_{\rm
  r}$/d.o.f. =$0.66/13$. We estimated two parameters from the fit: the
temperature of the plasma $kT=2.44\pm0.40$~keV and the unabsorbed
$3-8$~keV flux $(8.70\pm0.70)\times10^{-13}$\,\ergscm.

The emission of the Arches cluster extended emission was approximated
with power-law model and a Gaussian line with position and width fixed
at 6.4 keV and 0.1~keV, respectively. The model gives acceptable fit
statistics $\chi^{2}_{\rm r}$/d.o.f.=$1.02/12$ and allows for a
constraint on the power-law slope $\Gamma=2.06\pm0.26$ and the
unabsorbed $3-20$~keV flux $F_{\rm
  3-20}=(9.33\pm1.34)\times10^{-13}$\,\ergscm. The total flux of the
6.4~keV Gaussian line was estimated to be $(1.38\pm0.50)\times10^{-5}$
photons~cm$^{-2}$~s$^{-1}$.

\section{Summary}

In this work we presented details of the 2D image analysis approach
used to decouple spatially confused emission components in the Arches
cluster complex \cite{krivonos2017}, namely, thermal emission of the
stellar cluster and surrounding extended non-thermal emission of the
molecular cloud. We demonstrated the capability of the method to
decouple spatial emission components and validated its reliability in
comparison with standard approach.

\section*{Acknowledgments}
RK acknowledges support from the Russian
Basic Research Foundation (grant 16-02-00294), the Academy of Finland
(grant 300005) and hospitality of the Tuorla Observatory.

\begin{multicols}{2}

\end{multicols}

\end{document}